\documentclass[useAMS,usenatbib]{mn2e}
\usepackage{graphicx,fleqn,fix2col,rotating}
\DeclareGraphicsExtensions{.eps,.ps,.eps.gz,.ps.gz,.eps.Z}
\title[A J-band detection of the donor star in OY Car]{A J-band detection of the donor star in the dwarf nova OY Carinae, and an optical detection of its `iron curtain'}

\author[C.M.~Copperwheat et al.]{C.M.~Copperwheat$^{1}$, T.R.~Marsh$^{1}$, S.G.~Parsons$^{1}$, R.~Hickman$^{1}$\newauthor D.~Steeghs$^{1}$, E.~Breedt$^{1}$, V.S.~Dhillon$^{2}$, S.P.~Littlefair$^{2}$ and C.~Savoury$^{2}$\\\\
$^{1}$ Department of Physics, University of Warwick, Coventry, CV4 7AL, UK\\
$^{2}$ Department of Physics and Astronomy, University of Sheffield, Sheffield, S3 7RH, UK\\
}

\date{Received: }

\begin{document}

\newcommand{\dg} {^{\circ}}
\outer\def\gtae {$\buildrel {\lower3pt\hbox{$>$}} \over
{\lower2pt\hbox{$\sim$}} $}
\outer\def\ltae {$\buildrel {\lower3pt\hbox{$<$}} \over
{\lower2pt\hbox{$\sim$}} $}
\newcommand{\ergscm} {erg s$^{-1}$ cm$^{-2}$}
\newcommand{\ergss} {erg s$^{-1}$}
\newcommand{\ergsd} {erg s$^{-1}$ $d^{2}_{100}$}
\newcommand{\pcmsq} {cm$^{-2}$}
\newcommand{\ros} {{\it ROSAT}}
\newcommand{\xmm} {\mbox{{\it XMM-Newton}}}
\newcommand{\exo} {{\it EXOSAT}}
\newcommand{\sax} {{\it BeppoSAX}}
\newcommand{\chandra} {{\it Chandra}}
\newcommand{\hst} {{\it HST}}
\newcommand{\subaru} {{\it Subaru}}
\def\rchi{{${\chi}_{\nu}^{2}$}}
\newcommand{\Msun} {$M_{\odot}$}
\newcommand{\Mwd} {$M_{wd}$}
\newcommand{\Mbh} {$M_{\bullet}$}
\newcommand{\Lsun} {$L_{\odot}$}
\newcommand{\Rsun} {$R_{\odot}$}
\newcommand{\Zsun} {$Z_{\odot}$}
\newcommand{\Mjup} {$M_{J}$}
\newcommand{\Rjup} {$R_{J}$}

\def\Mdot{\hbox{$\dot M$}}
\def\mdot{\hbox{$\dot m$}}
\def\mincir{\raise -2.truept\hbox{\rlap{\hbox{$\sim$}}\raise5.truept
\hbox{$<$}\ }}
\def\magcir{\raise -4.truept\hbox{\rlap{\hbox{$\sim$}}\raise5.truept
\hbox{$>$}\ }}
\newcommand{\mnras} {MNRAS}
\newcommand{\aap} {A\&A}
\newcommand{\apj} {ApJ}
\newcommand{\apjl} {ApJL}
\newcommand{\apjs} {ApJS}
\newcommand{\aj} {AJ}
\newcommand{\pasp} {PASP}
\newcommand{\aaps} {AAPS}
\newcommand{\apss} {Ap\&SS}
\newcommand{\araa} {ARAA}
\newcommand{\nat} {Nature}
\newcommand{\pasj} {PASJ}
\newcommand{\actaa} {ACTA}
\newcommand{\ha}{\hbox{$\hbox{H}\alpha$}}
\newcommand{\hb}{\hbox{$\hbox{H}\beta$}}
\newcommand{\hg}{\hbox{$\hbox{H}\gamma$}}
\newcommand{\heii}{\hbox{$\hbox{He\,{\sc ii}\,$\lambda$4686\,\AA}$}}
\newcommand{\hei}{\hbox{$\hbox{He\,{\sc i}\,$\lambda$4472\,\AA}$}}

\maketitle

\begin{abstract} 
Purely photometric models can be used to determine the binary parameters of eclipsing cataclysmic variables with a high degree of precision. However, the photometric method relies on a number of assumptions, and to date there have been very few independent checks of this method in the literature. We present time-resolved spectroscopy of the $P=90.9$ min eclipsing cataclysmic variable OY Carinae obtained with X-shooter on the VLT, in which we detect the donor star from K~{\sevensize I} lines in the J-band. We measure the radial velocity amplitude of the donor star $K_2 = 470.0 \pm 2.7$ km/s, consistent with predictions based upon the photometric method ($470 \pm 7$km/s). Additionally, the spectra obtained in the UVB arm of X-shooter show a series of Fe~{\sevensize I} and Fe~{\sevensize II} lines with a phase and velocity consistent with an origin in the accretion disc. This is the first unambiguous detection at optical wavelengths of the `iron curtain' of disc material which has been previously reported to veil the white dwarf in this system. The velocities of these lines do not track the white dwarf, reflecting a distortion of the outer disc that we see also in Doppler images. This is evidence for considerable radial motion in the outer disk, at up to $90$km/s towards and away from the white dwarf.

\end{abstract}

\section{INTRODUCTION}  
\label{sec:intro}

The study of close binary stars is driven by the need to understand the numbers and properties of the many possible outcomes of binary star evolution. This is of interest in many astrophysical contexts, such as the need to determine the properties of Type Ia supernovae progenitors as a function of redshift, and the detection of gravitational waves, for which close binaries are predicted to be among the strongest sources \citep{Nelemans04}. A useful population are the cataclysmic variables (CVs: \citealt{Warner95}). These close binaries consist of a white dwarf accreting matter from a main sequence donor star, and provide us with a large, homogeneous and easily observed population which can be compared to theoretical predictions.

The binary parameters of eclipsing systems can be determined with high precision via a purely photometric method \citep{Bailey79,Smak79,Cook84,Wood86} in which the eclipse features in the light curve are used to infer the geometry of the system. The number of known eclipsing systems has increased significantly in recent years thanks to programmes such as the Sloan Digital Sky Survey (SDSS: \citealt{Szkody09}), and so greater reliance has been placed on this photometric technique (see, e.g., \citealt{Littlefair08, Copperwheat10, Copperwheat11, Savoury11}). The photometric method is based on two key assumptions: first, it is assumed that we see the bare white dwarf and can therefore measure its radius, and hence its mass, using models of white dwarf mass-radius relations. Secondly, it is assumed that the bright-spot where the gas stream hits the accretion disc lies on the ballistic trajectory of the gas stream, which follows a path set by the binary mass ratio. To date there have been very few independent checks of the photometric method. \citet{Wade88} presented a radial velocity study of the secondary star in Z~Cha, and found a value for the donor star orbital velocity $K_2 = 430 \pm 16$km/s. In comparison, the photometric study of \citet{Wood86} found $K_2 = 371.0 \pm 2.7$km/s, a discrepancy of more than $3\sigma$. This was partly attributed to an overestimation of the white dwarf radius in \citet{Wood86}, but later work accounting for this (increasing the photometric determination of $K_2$ to $389$ -- $406$km/s) was still $\sim$$2\sigma$ different from the spectroscopic finding \citep{Wood90}. More recent spectroscopic studies (IP Peg, \citealt{Watson03}; CTCV J1300-3052, \citealt{Savoury11a}) have shown good agreement with photometric results. However, the small number of test cases to date is concerning, given the ubiquity of the photometric method. There have been some attempts to verify the method by inferring the white dwarf orbital velocity ($K_1$) from the velocities of the emission lines (see, e.g. \citealt{Tulloch09}). However, these measurements are generally based on the assumption that the accretion disc is perfectly symmetric around the white dwarf. This is often not the case, and while there have been attempts to correct for a non-symmetrical disc \citep{Marsh88a} we would still consider these $K_1$ determinations to be less reliable than measurements of the donor star.

The capabilities of modern spectrographs far exceed the instrument used by \citet{Wade88}. In particular, the wavelength range and throughput of X-shooter mounted on the ESO Very Large Telescope (VLT) enables us to determine a value of $K_2$ with a precision comparable to that which can be currently achieved via the photometric method. We therefore chose to measure this orbital velocity for the donor star in a CV and thus provide a more precise test of the light curve fitting method. The system we selected was OY Carinae (henceforth OY Car), a bright and short-period ($P=90.9$ min) eclipsing system with a prominent white dwarf eclipse. The best photometric determinations for the binary parameters in OY Car to date were published in \citet{Littlefair08}.

\section{OBSERVATIONS AND REDUCTION}
\label{sec:obs}

We observed OY Car for a total of $1.8$h on the night of $2010$ February $10$ with X-shooter \citep{Dodorico06} mounted on VLT UT2 ({\it Kueyen}). The X-shooter spectrograph consists of 3 independent arms (UVB, VIS and NIR), giving a simultaneous wavelength coverage from $3000$ to $24800$\AA. We obtained $28$ consecutive spectra of $185$, $180$ and $205$s in length for the UVB, VIS and NIR arms respectively. The cycle time per exposure was $\sim$$235$s. We binned by a factor of two both spatially and in the dispersion direction for the UVB and VIS arms, and used slit widths of $1.0$'', $1.2$'' and $0.9$'' for the UVB, VIS and NIR arms respectively, resulting in FWHM spectral resolutions of $0.7$, $1.0$ and $2.9$\AA. Weather conditions were excellent, with sub-arcsecond seeing and photometric transparency throughout.

\begin{figure*}
\centering
\includegraphics[angle=270,width=1.0\textwidth]{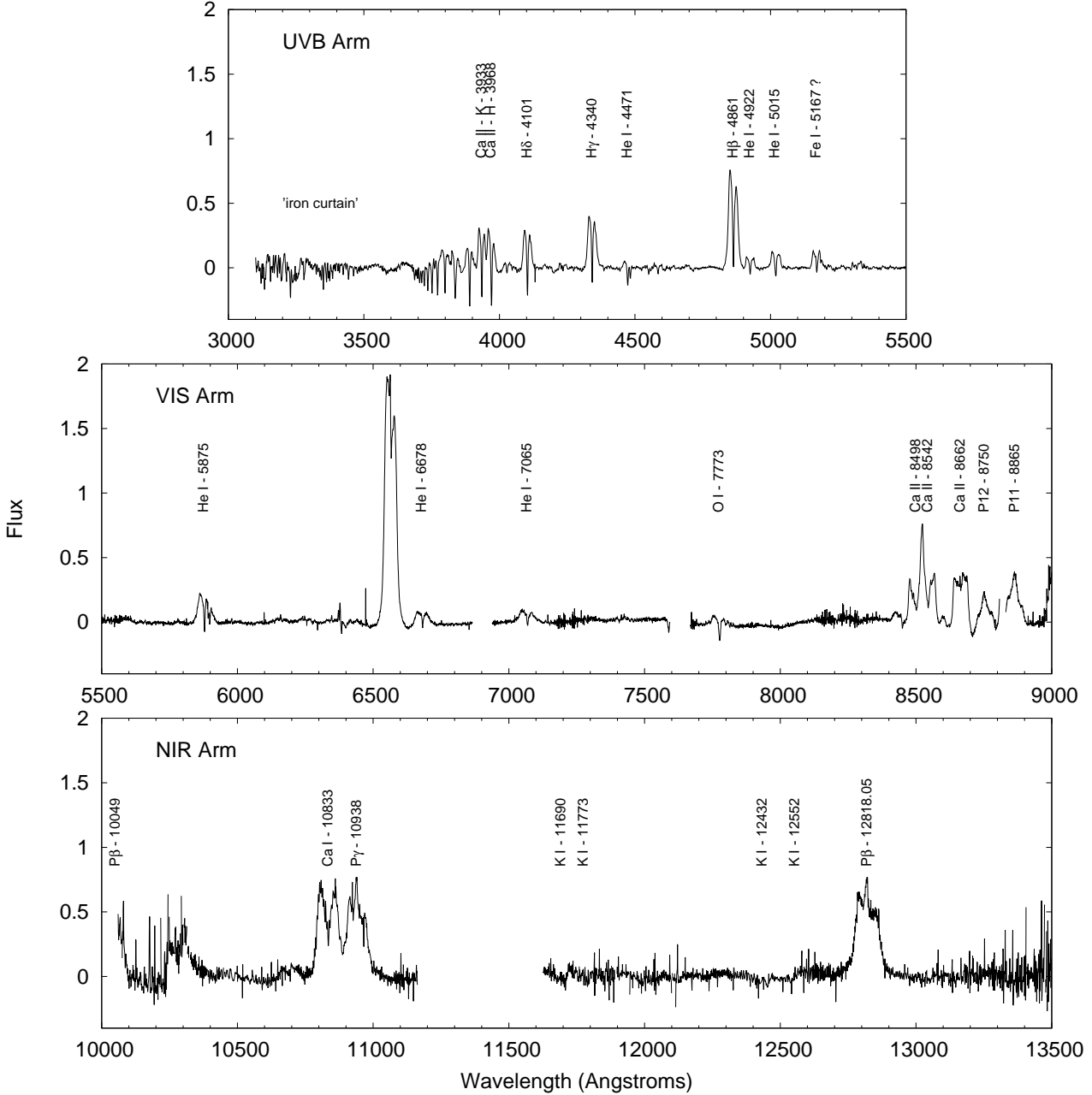}
\hfill
\caption{Averaged and continuum subtracted spectra. The top, middle and bottom panels show the spectrum obtained in the UVB, VIS and NIR arms of the spectrograph, respectively. We give the wavelength in Angstroms, and the panels are of different widths so as to keep the scale consistent. We identify the most prominent features, and mask out the regions of the spectrum which are most contaminated by telluric features.} \label{fig:average} \end{figure*}

We reduced these data using version $1.3.7$ of the X-shooter pipeline. The standard recipes were used to optimally extract and wavelength calibrate each spectrum. We removed the instrumental response using observations of the spectrophotometric standard star GD 71. One complication was that, owing to an oversight, the data were obtained in `stare' mode, rather than by nodding the object along the slit, as is typical for long slit infrared spectroscopy. The result is the sky subtraction is poorer than is usually possible with X-shooter. We divided our spectra by the spectrum of a telluric standard star taken at a similar airmass to reduce telluric absorption. Fortunately, most of the spectral lines we use are displaced from the worst telluric features; the optical data are not affected. We plot the averaged and continuum subtracted spectra from all three arms in Figure~\ref{fig:average}.

\section{RESULTS}
\label{sec:results}

\subsection{The orbital velocity of the donor star}
\label{sec:nir}

\begin{figure*}
\centering
\includegraphics[angle=270,width=1.0\textwidth]{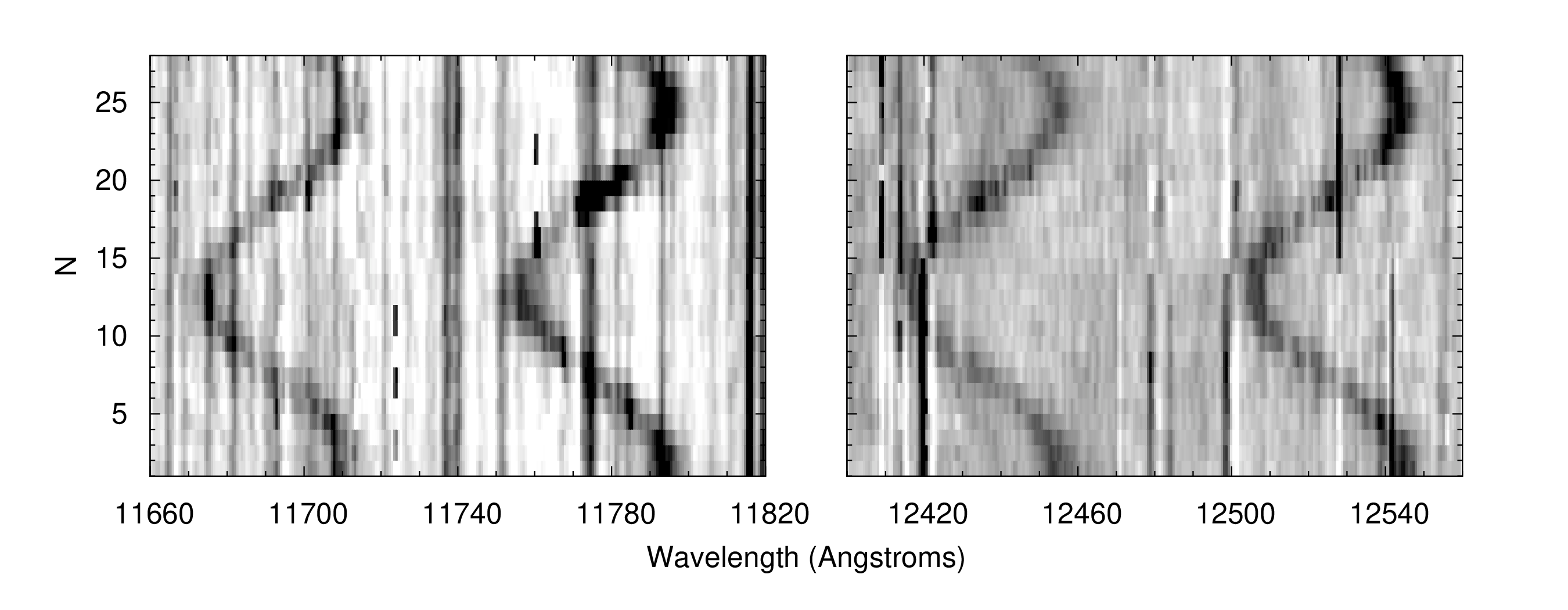}
\hfill
\caption{Trailed spectra obtained in the NIR arm. We plot the regions around the  $\lambda \lambda$ $11692$, $11778$\AA \ and  $\lambda \lambda$ $12437$, $12528$\AA \  K~{\sevensize I} doublets. These absorption lines move in phase with the M-dwarf donor. The scale is such that the continuum is white with darker absorption features. `$N$' denotes the number of the spectrum. The $N=19$ spectrum was obtained around phase $0$.} \label{fig:nir_trail} \end{figure*}

In order to measure $K_2$ in Z~Cha, \citet{Wade88} used the $\lambda \lambda$ $8183$, $8194$ \AA \ Na~{\sevensize I} doublet to track the motion of the donor star. The wavelength coverage and sensitivity of X-shooter means other absorption lines are available, and infrared spectra of M, L and T dwarfs show promising features in the J-band (\citealt{Cushing05}, table 6) which have been used in previous CV studies to detect the donor star \citep{Littlefair00}. We find the K~{\sevensize I} doublets at $\lambda \lambda$ $11690$, $11772$ \AA \ and $\lambda \lambda$ $12435$, $12522$ \AA \ to be much stronger features in our OY Car data compared to the Na~{\sevensize I} doublet, and we used these features to track the donor star. We plot the trailed spectra around these absorption lines in Figure \ref{fig:nir_trail}. Some residual telluric contamination is apparent in these trailed spectra: the $12435$\AA \ line is particularly affected. We masked out these sections of data when applying our model fits.

\begin{figure}
\centering
\includegraphics[angle=270,width=1.0\columnwidth]{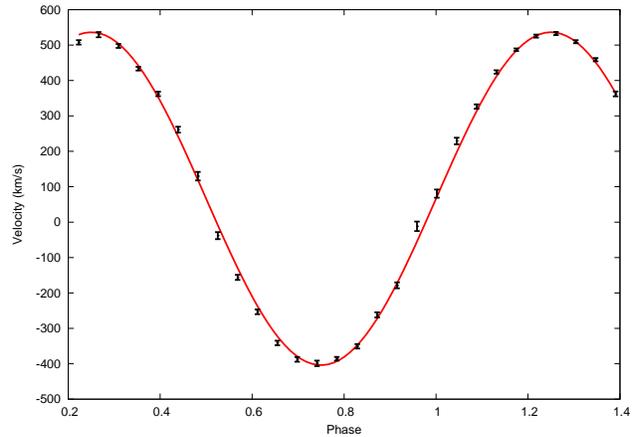}
\hfill
\caption{Radial velocity curve for the secondary star. The solid line is the sinusoid corresponding to our combined fit orbital solution, as determined in Section \ref{sec:nir}.} \label{fig:nir_indiv} \end{figure}

We fitted the spectra with a model consisting of a straight line and Gaussians for the K absorption lines, allowing the position of each Gaussian to change velocity according to
\begin{eqnarray}
V = \gamma + K_2 \sin(2\pi\phi), \nonumber
\end{eqnarray}
\noindent where $\gamma$ is the offset of the line from its rest wavelength and $\phi$ is the orbital phase of the spectrum. We obtained the rest wavelengths from the {\it NIST} Atomic Spectra Database \citep{Ralchenko11}. The spectra were fitted simultaneously using a Levenberg-Marquardt minimisation \citep{Press02}, and the results are listed in Table \ref{tab:nirfit}. The zero point for the orbital phase was a free parameter in these fits, and we find it to be at HJD $=2443993.54892(4)$. We used the orbital period from the quadratic ephemeris given in \citet{Greenhill06}.  We give the results for each line fitted individually with a single Gaussian model, as well as a combined fit in which the four lines are fitted together with a common $\gamma$, $K_2$ and velocity width. The combined fit gives a value of $K_2 = 470.1 \pm 2.0$km/s for the orbital velocity of the donor star. The model fits typically give a reduced $\chi^2$ of $\sim$$1.4$, so we rescale
the uncertainties to give a reduced $\chi^2$ of $1$, and it is these values we quote in the table. We also make a second determination of the combined fit, for which we use the bootstrap method \citep{Efron79, Efron93}. In this method we determine the parameter values and uncertainties via fitting a large number of individual sets of spectra which have been resampled from the original set. The two combined fits give parameter values which are in excellent agreement, with the uncertainties slightly higher for results determined via the bootstrap method. We used the combined fit as a starting point to fit each spectrum individually, and plot the resultant radial velocity curve in Figure \ref{fig:nir_indiv} along with the sinusoid corresponding to our orbital solution. The sinusoid is a good fit to the radial velocity -- there is little to no evidence for irradiation as is observed in other CVs (e.g. EX~Dra, \citealt{Billington96}). 

\begin{table}
\caption{Parameter fits to the four K~{\sevensize I} lines observed in the NIR arm. We list the systemic velocity $\gamma$, the donor star orbital velocity $K_2$, and the full-width half maximum (FWHM) of the absorption line, which we also express in velocity units. We list the results from fitting each line individually with a single Gaussian model. The uncertainties quoted are the formal errors, scaled to give a reduced $\chi^2$ of $1$. We also list the results of combined fits obtained with a model consisting of four Gaussians. We give two combined results: one in which the uncertainties are the scaled formal errors, as for the individual fits, and a second in which we determined the parameter value and uncertainty via a bootstrap method. The bootstrap method gives slightly larger uncertainties.}
\label{tab:nirfit}
\begin{center}
\begin{tabular}{lr@{\,$\pm$\,}lr@{\,$\pm$\,}lr@{\,$\pm$\,}l}
\multicolumn{1}{c}{Wavelength}     &\multicolumn{2}{c}{$\gamma$} &\multicolumn{2}{c}{$K_2$} &\multicolumn{2}{c}{FWHM }\\
\multicolumn{1}{c}{(\AA)}&\multicolumn{2}{c}{(km/s)}&\multicolumn{2}{c}{(km/s)}&\multicolumn{2}{c}{(km/s)}\\
\hline
$11690.22$                 &$74.5$ &$4.6$      &$457.1$ &$6.2$     &$235.6$ &$10.8$\\
$11772.84$                 &$58.8$ &$2.7$      &$473.2$ &$3.4$     &$246.6$ &$6.4$\\
$12432.27$                 &$77.4$ &$4.3$      &$464.1$ &$5.6$     &$243.7$ &$8.5$\\
$12522.14$                 &$67.4$ &$3.0$      &$470.5$ &$4.1$     &$206.9$ &$6.7$\\
Combined ($\chi^2$=1)      &$66.2$ &$1.6$      &$470.1$ &$2.0$     &$238.9$ &$3.8$\\
Combined (Bootstrap)       &$66.2$ &$2.5$      &$470.0$ &$2.7$     &$239.1$ &$4.4$\\

\hline
\end{tabular}
\end{center}
\end{table}

The combined (bootstrap) fit gives a value of $K_2 = 470.0 \pm 2.7$km/s for the orbital velocity of the donor star. When we fit the lines individually we find consistent values for $K_2$ for three of the four lines, with the $11690$\AA \ value around $1.5\sigma$ lower than the value for the combined fit. Our determination of the systemic velocity $\gamma$ from the combined fit is $66.2 \pm 2.5$km/s, although this determination is less secure since three of individual fits result in $\gamma$ values which differ significantly from the combined value. The formal errors we quote would seem to be an underestimation of the true uncertainty in this parameter at least, probably due to the telluric absorption.

\citet{Wood90} showed that the photometrically determined value of $K_2$ depends upon the white dwarf radius, the determination of which depends on the limb darkening coefficient used. For OY~Car they reported $K_2$ values of $470 \pm 8$ and $455 \pm 9$km/s for a limb darkening coefficient of $u_{WD}=0$ or $1$ respectively, where the white dwarf intensity distribution $I(\theta) = I_0(1 + u_{WD} - u_{WD} \cos\theta)$. However, they used a zero temperature mass-radius relationship for the white dwarf, which introduced a significant bias. \citet{Littlefair08} revisited the system parameters of OY Car, and improved on the \citet{Wood90} determination by using a more realistic finite temperature model for the white dwarf. They assumed $u_{WD} = 0.5$ and $T_{WD} = 16,5000$K \citep{Horne94} and find $K_2 = 470 \pm 7$km/s. This is in excellent agreement with our spectroscopic value of $470.0 \pm 2.7$km/s.

\subsection{Rotational broadening of the donor star}
\label{sec:rotation}

\begin{figure}
\includegraphics[angle=270,width=1.0\columnwidth]{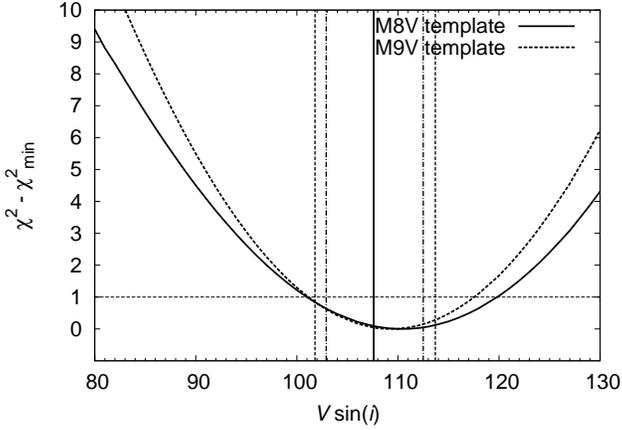}
\caption{Curves of reduced $\chi^2$ versus $V_{rot}\sin i$, obtained via the rotational broadening measurement procedure described in Section \ref{sec:rotation}. For the template star we use LHS~2021, an M8 dwarf (solid curve) and LHS~2065, an M9 dwarf (dashed curve). We assume a limb darkening coefficient of $0.6$. The solid vertical line marks the value of $v\sin i$ determined using the \citet{Wood90} binary parameters. The vertical dashed lines mark the $1\sigma$ error on this value, using the \citet{Littlefair08} parameter uncertainties. The vertical dot-dashed lines show the uncertainty when we use our tighter constraint on $K_2$.} \label{fig:broaden} \end{figure}

As an additional test of the photometric parameters, we measured the rotational broadening of the donor star $V_{rot}$ following the method detailed in \citet{Marsh94}. Using the model fit detailed in Section \ref{sec:nir}, we applied an offset to each spectrum to remove the orbital variation and then averaged the spectra. We then applied an artificial rotational broadening to the spectrum of a template star, and subtracted the result, multiplied by a constant representing the fraction of flux in the donor star, from the averaged OY Car spectrum. The constant was varied to optimise the subtraction, and the $\chi^2$ difference between the residual spectrum and a smoothed version of itself is computed. This process was repeated for a range of rotational broadenings. By comparing our spectra with the spectroscopic sequence presented in \citet{Cushing05} we identify the donor in OY Car to be a late M-dwarf (M8/M9) or possibly an early T-dwarf. We rule out an earlier type M-dwarf since in these stars the K~{\sevensize I} features are weaker, and there are a number of Fe absorption lines of comparable strength around the $\lambda \lambda$ $11690$, $11772$ \AA \ doublet which we do not observe in our OY Car spectra. We therefore used used LHS~2021 and LHS~2065 as our template stars, which are of spectral type M8 and M9 respectively. An additional input parameter is the limb darkening coefficient for the donor star. \citet{Claret98} listed limb darkening coefficients for low mass stars, and from these tables we find a value of $u_{2} \sim 0.6$ -- $0.7$ is an appropriate $J$-band coefficient for the donor. We plot in Figure \ref{fig:broaden} the resulting $\chi^2$ curves for $u_2 = 0.6$. The minima of these curves give our preferred values of $V_{rot}\sin i$, which we find to be $110$km/s and $109$km/s for the M8 and M9 templates, respectively. 

The quantity $V_{rot}\sin i$ is related to the mass ratio $q$ and $K_2$ via
\begin{eqnarray}
V_{rot}\sin i = K_2 (q+1) {R_2 \over a} \nonumber
\end{eqnarray}
\citep{Horne86}, where for a Roche-lobe filling star $R_2/a$ is a function of $q$ \citep{Eggleton83}. Using the photometrically determined parameters from \citet{Littlefair08}, we find $V_{rot}\sin i = 108 \pm 6$km/s. This uncertainty decreases to $\pm 5$km/s when we use our spectroscopically determined constraint on $K_2$. We indicate this photometric determination of $V_{rot}\sin i$ in Figure \ref{fig:broaden}. The figure demonstrates that, for both template stars, our spectroscopically determined value of $V_{rot}\sin i$ for $u_{2}=0.6$ is consistent with the photometric value to within the uncertainties. We should note that this $u_{2}$ is appropriate for the continuum of the donor star in the $J$-band, but the coefficient for the absorption lines may be somewhat different, which would change the optimum spectroscopic value of $V_{rot}\sin i$. However, we find changing $u_2$ in increments of $0.1$ shifts the optimum value by only $1$ -- $2$km/s, and so given the uncertainty in the photometric determination of $V_{rot} \sin i$, the spectroscopic value will still be consistent even if the coefficient for the lines is very different to our assumed value.

\subsection{Doppler maps: emission from the disc and donor star}
\label{sec:dopp}

\begin{figure*}
\centering
\includegraphics[angle=270,width=1.0\textwidth]{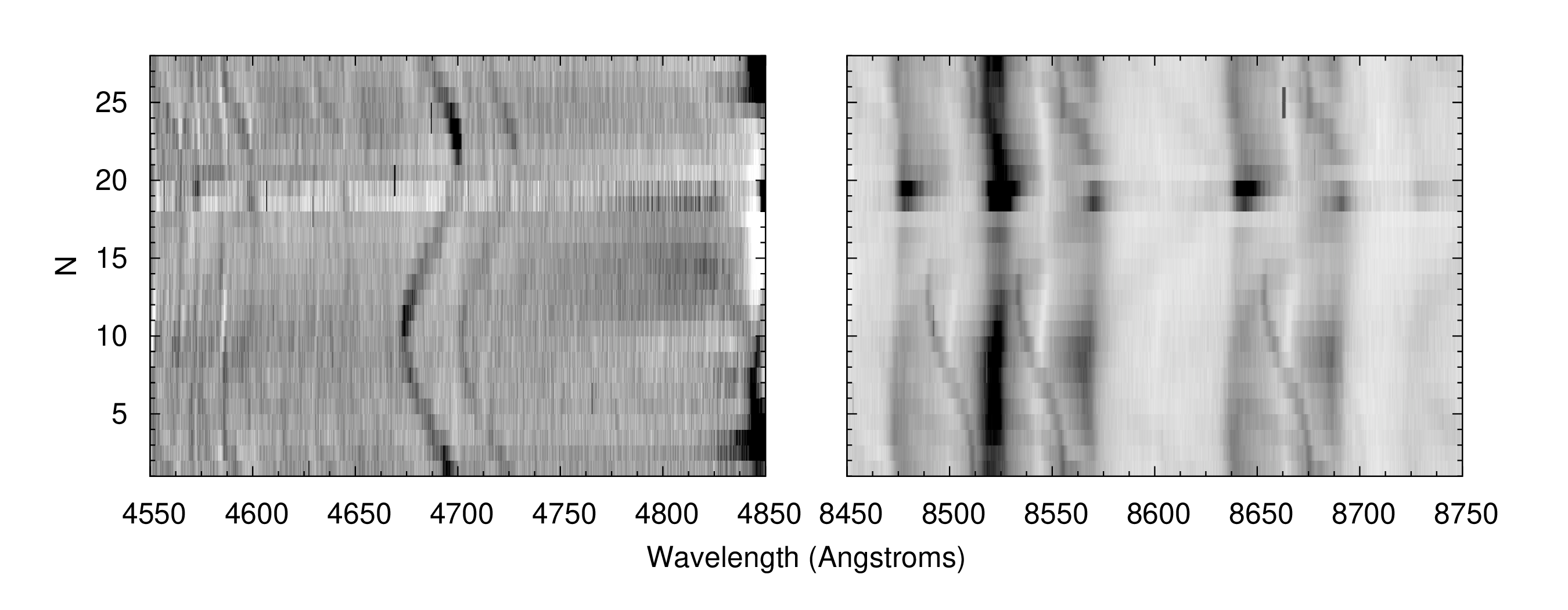}
\includegraphics[angle=0,width=1.0\textwidth]{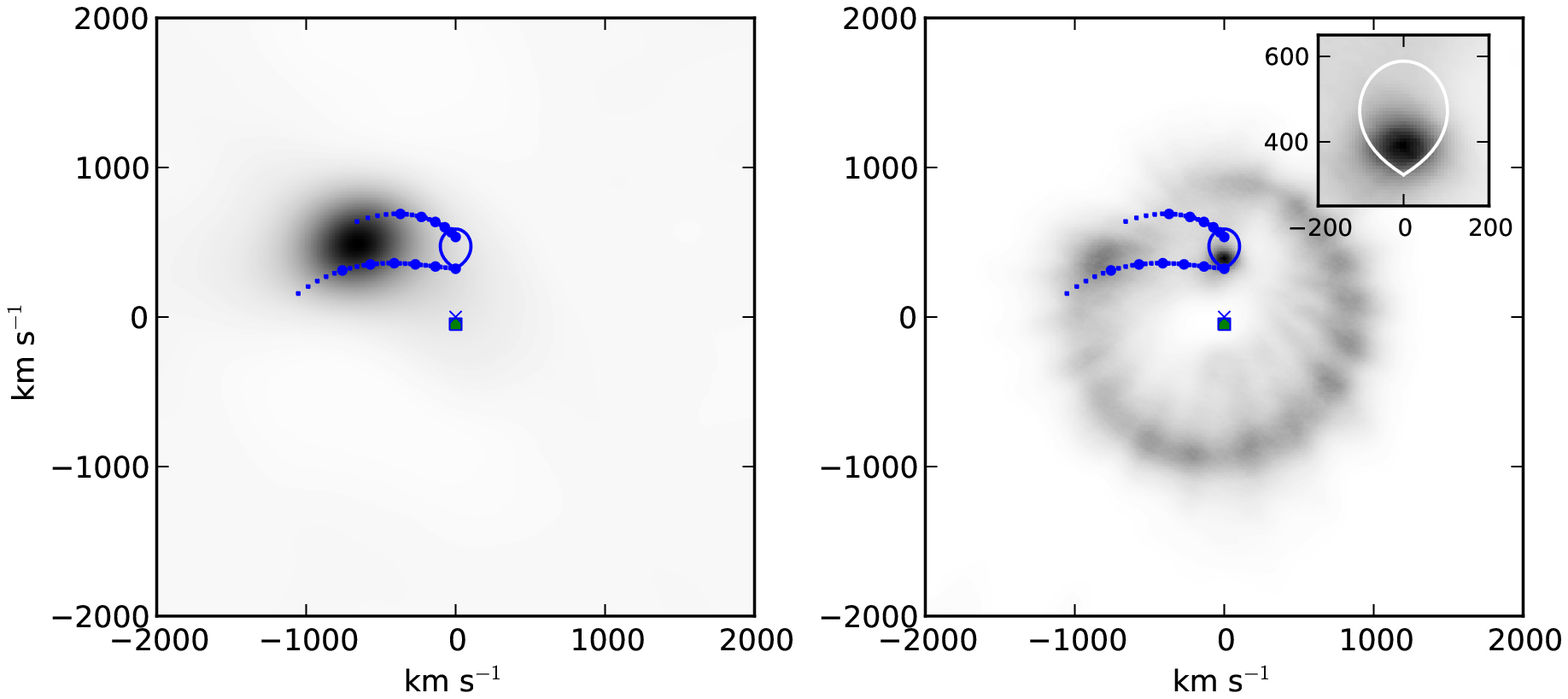}
\hfill
\caption{{\bf Top}: trailed spectra around the $4686$\AA \  He~{\sevensize II} emission line (left) and the $\lambda \lambda$ $8498$, $8542$, $8662$\AA Ca~{\sevensize II} triplet (right). The $4713$\AA \ He{\sevensize I} line is also visible in the He~{\sevensize II} trail. The scale is reversed for these plots compared to the other trailed spectra in this paper, with the dark lines showing emission features rather than absorption. `$N$' denotes the number of the spectrum. The $N=19$ spectrum was obtained around phase $0$. {\bf Bottom}: Doppler maps for the He~{\sevensize II} (left) and Ca~{\sevensize II} (right) emission lines. In the Ca~{\sevensize II} we produce a combined map from the three lines of the triplet. The `spoking' evident in this map is an artefact due to the relatively low phase resolution of our spectra. We overplot in blue the Roche lobe of the donor star and the streams for a model with $q=0.102$, $K_1=48$ and $K_2=470$ \citep{Wood90}. The streams denote the velocity of material in the gas stream itself (lower) and the velocity of the disc along the stream (upper). We mark with a cross the centre of mass, and with a square the position of the white dwarf. Points of equal distance along the two streams are marked, in units of the distance to the inner Lagrangian point from the centre of mass of the white dwarf. The small and large points mark increments of $0.02$ and $0.1$ of this distance, respectively.} \label{fig:dm} \end{figure*}

We create Doppler maps using a Maximum Entropy Method ({\sevensize MEM}, \citealt{Marsh88, Marsh01}) for the $4686$\AA \  He~{\sevensize II} emission line, and the Ca~{\sevensize II} triplet at $\lambda \lambda$ $8498$, $8542$, $8662$\AA. We plot the trailed spectra and Doppler maps in Figure \ref{fig:dm}. The Ca~{\sevensize II} map is a single, common map to represent all three lines, with a relative scaling of $1.00$, $1.25$ and $1.06$ for the three lines respectively. If we examine the Ca~{\sevensize II} map first, we see a ring of emission which originates in the accretion disc. The velocity of this emission is $\sim$$900$km/s, which is towards the high end of the range observed in other CVs (e.g. $600$km/s in U~Gem, \citealt{Marsh90}; see also many other examples in \citealt{Kaitchuck94}). This implies a small accretion disc, and is consistent with the short orbital period and large white dwarf mass of OY Car. We also see the `bright spot' of emission where the gas stream impacts the accretion disc, as well as Ca~{\sevensize II} emission from the donor star. This donor emission can be identified as a sharp S-wave component in the trailed spectra, and in the Doppler map we see that it is concentrated in the irradiated face of the star. We also created (but do not plot) Doppler maps of the Balmer lines. These show the same features which we observe in the Ca~{\sevensize II} map, but the donor emission is not as sharply defined.

Turning to the He~{\sevensize II} emission, in the trailed spectra we see an S-wave component with an amplitude and phase that is inconsistent with a donor star origin. In the Doppler map this emission resolves to a single component in the bright spot. Both the Ca~{\sevensize II} and He~{\sevensize II} bright spot emission have kinematics which place them between the disc and stream kinematics. The emission in both maps is quite smeared out, but the Ca~{\sevensize II} emission may be slightly closer to the velocity of the incoming gas stream material. This would suggest that the emission in He~{\sevensize II} is from the mixture of gas stream and disc material further downstream, as is seen in other CVs (e.g. U Gem, \citealt{Marsh90}). 

\subsection{The `iron curtain' in the accretion disc}
\label{sec:uvb}

\begin{figure*}
\centering
\hspace{-2.8cm}
\includegraphics[angle=270,width=1.1\textwidth]{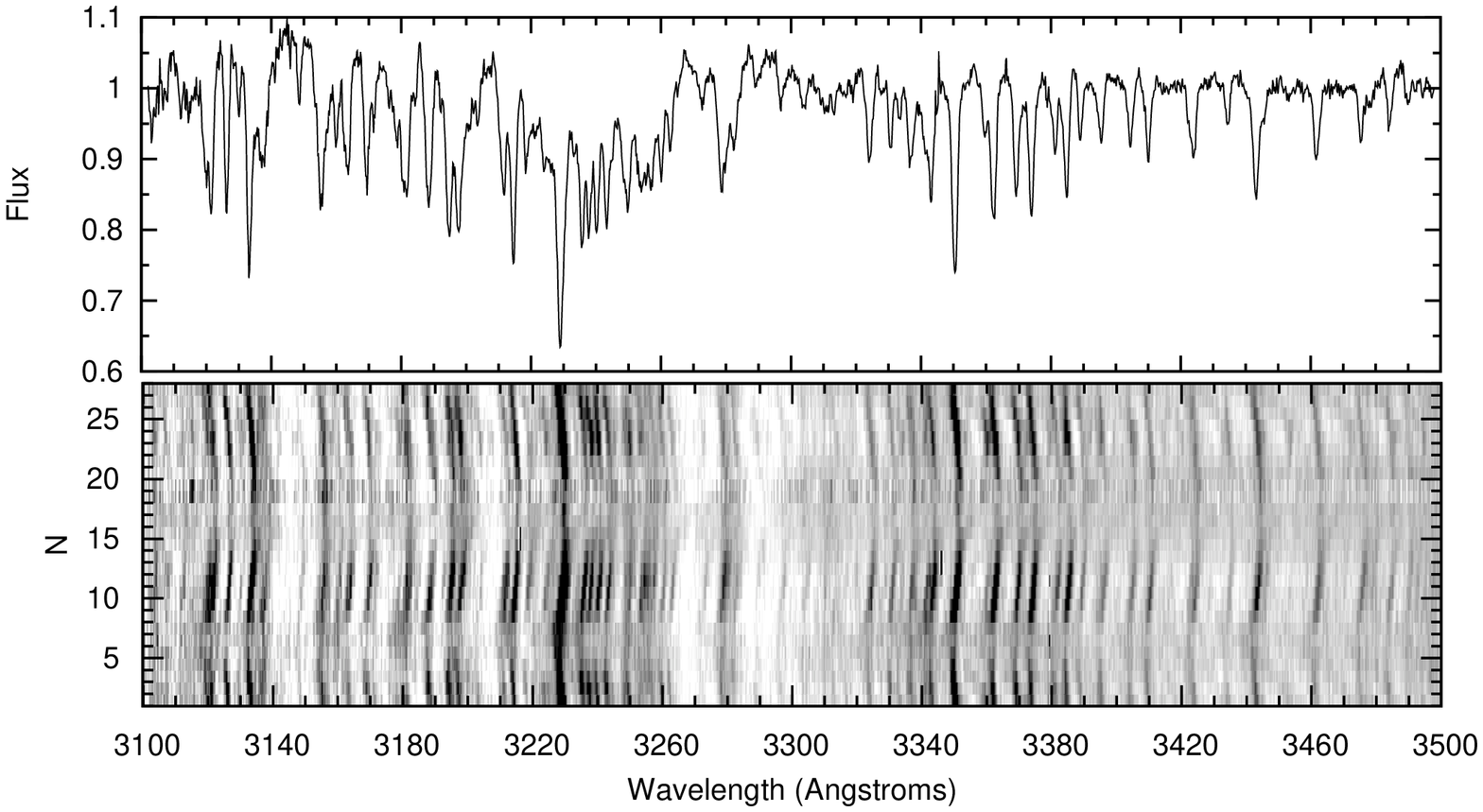}
\vspace{-2.5cm}
\caption{{\bf Bottom}: Trailed spectra obtained in the UVB arm between $3100$ and $3500$\AA. We observe a series of absorption lines, with a phase and velocity consistent with an accretion disc origin. The scale is such that the continuum is white with darker absorption features. `$N$' denotes the number of the spectrum. The $N=19$ spectrum was obtained around phase $0$. {\bf Top}: Averaged spectrum over the same wavelength range, with the velocity variations subtracted.} \label{fig:uvb_trail} \end{figure*}

In the UVB arm ($3000$ -- $5600$\AA), we observe many absorption lines with a common phase and velocity. The velocity is such that these features cannot be attributed to either the white dwarf or the donor star, and so we conclude they originate in the accretion disc. The majority of these features lie in a wavelength range of $3100$ -- $3500$\AA \ (Figure \ref{fig:uvb_trail}), but additional lines can be observed longwards of this to $\sim$$5000$\AA.

\begin{table}
\caption{Absorption features identified in the spectra obtained in the UVB arm. The rest wavelengths are obtained from the {\it NIST} Atomic Spectra Database \citep{Ralchenko11}. }
\label{tab:ironlines}
\begin{center}
\begin{tabular}{lclc}
Line    &Wavelength &Line    &Wavelength\\
        &(\AA)      &        &(\AA)\\
\hline
Fe~{\sevensize I}      &$3120.44$  &Fe~{\sevensize I}      &$3125.65$\\
Fe~{\sevensize I}      &$3129.33$  &Fe~{\sevensize I}      &$3133.52$\\
Fe~{\sevensize I}      &$3135.86$  &Fe~{\sevensize I}      &$3147.79$\\
Fe~{\sevensize II}     &$3154.20$  &Fe~{\sevensize II}     &$3162.80$\\
Fe~{\sevensize I}      &$3168.85$  &Fe~{\sevensize I}      &$3180.22$\\
Fe~{\sevensize II}     &$3187.30$  &Fe~{\sevensize I}      &$3193.30$\\
Fe~{\sevensize I}      &$3196.93$  &Fe~{\sevensize I}      &$3210.83$\\
Fe~{\sevensize II}     &$3213.31$  &Fe~{\sevensize I}      &$3217.38$\\
Fe~{\sevensize I}      &$3222.07$  &Fe~{\sevensize I}      &$3227.80$\\
Fe~{\sevensize I}      &$3234.61$  &Fe~{\sevensize I}      &$3236.22$\\
Fe~{\sevensize II}     &$3237.82$  &Fe~{\sevensize II}     &$3243.72$\\
Fe~{\sevensize I}      &$3248.20$  &Fe~{\sevensize I}      &$3259.99$\\
Fe~{\sevensize I}      &$3262.01$  &Fe~{\sevensize II}     &$3277.55$\\
Fe~{\sevensize II}     &$3281.29$  &Fe~{\sevensize I}      &$3323.74$\\
Fe~{\sevensize I}      &$3329.52$  &Fe~{\sevensize I}      &$3335.77$\\
Fe~{\sevensize I}      &$3342.29$  &Fe~{\sevensize I}      &$3349.72$\\
Fe~{\sevensize II}     &$3360.11$  &Fe~{\sevensize II}     &$3366.97$\\       
Fe~{\sevensize I}      &$3372.07$  &Fe~{\sevensize I}      &$3383.98$\\
Fe~{\sevensize II}     &$3388.13$  &Fe~{\sevensize I}      &$3394.58$\\
Fe~{\sevensize I}      &$3403.29$  &Fe~{\sevensize I}      &$3422.66$\\
Fe~{\sevensize I}      &$3433.57$  &Fe~{\sevensize I}      &$3442.36$\\
Fe~{\sevensize I}      &$3459.91$  &Ca~{\sevensize I}      &$3474.76$\\
Fe~{\sevensize I}      &$3483.01$  &Fe~{\sevensize I}      &$3565.38$\\
Fe~{\sevensize I}      &$3570.10$  &Fe~{\sevensize I}      &$3581.19$\\
Fe~{\sevensize I}      &$3618.77$  &Fe~{\sevensize I}      &$3631.46$\\
Fe~{\sevensize I}      &$3686.00$  &Fe~{\sevensize I}      &$3758.23$\\
Fe~{\sevensize I}      &$3761.41$  &Fe~{\sevensize I}      &$4045.81$\\
Fe~{\sevensize II}     &$4583.84$  &Fe~{\sevensize II}     &$4923.93$\\
Fe~{\sevensize II}     &$5018.43$  &\multicolumn{2}{c}{ }\\
\hline
\end{tabular}
\end{center}
\end{table}

\begin{figure}
\centering
\includegraphics[angle=270,width=1.0\columnwidth]{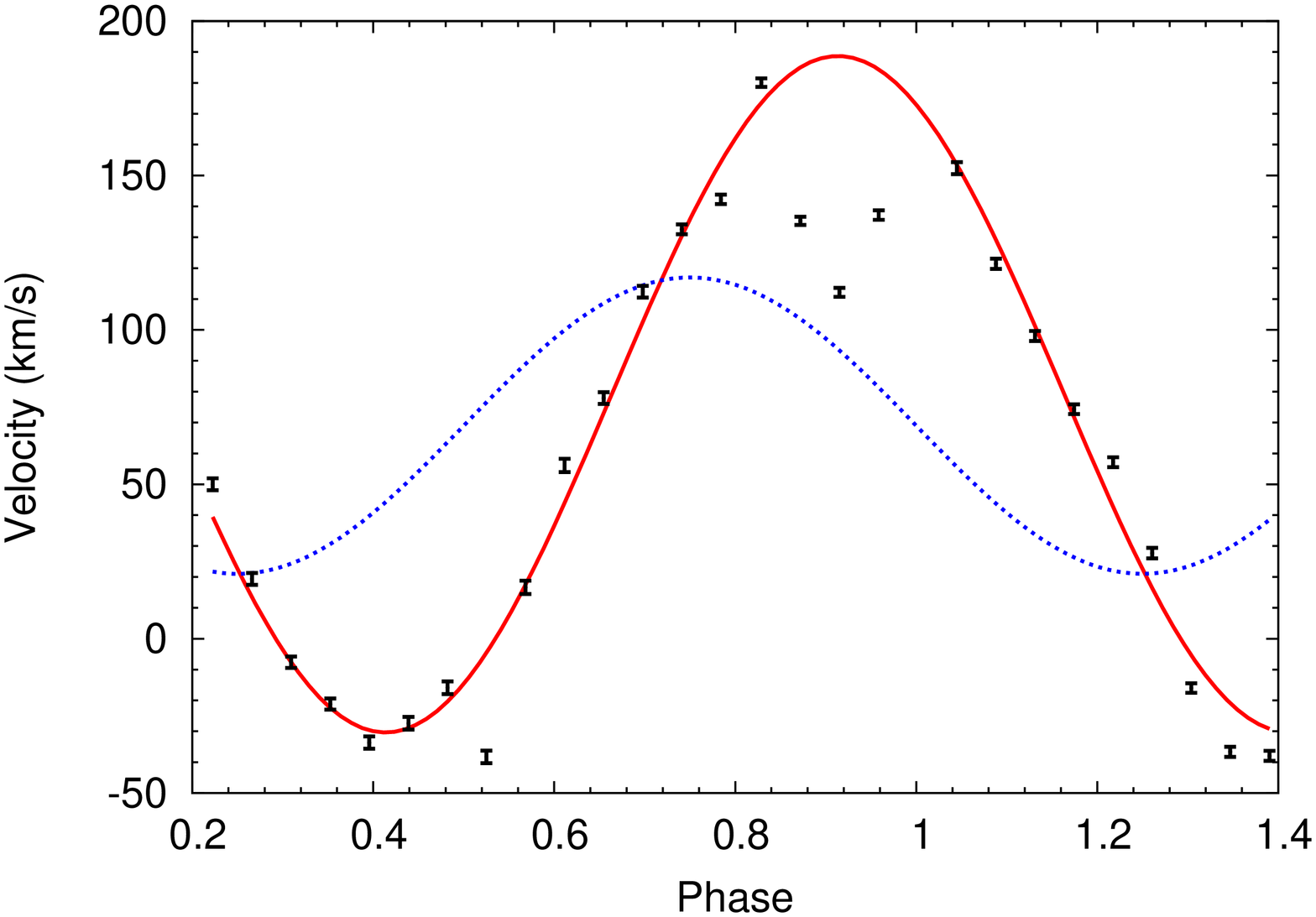}
\hfill
\caption{Radial velocity curve for the absorption lines observed in the UVB arm. The solid red line is the best sinusoid fit. The dashed blue line is the radial velocity curve for the white dwarf, assuming $K_1 = 48$km/s \citep{Wood90}.} \label{fig:uvb_indiv} \end{figure}

In Table \ref{tab:ironlines}, we list our identifications for these absorption lines. We find the majority to be Fe~{\sevensize I} and Fe~{\sevensize II} lines. These identifications, coupled with the evidence for an accretion disc origin, suggests we are seeing the `iron curtain' first identifed in this system by \citet{Horne94}. They obtained ultraviolet observations of OY Car with \hst, and found that the emission from the white dwarf was veiled by a forest of Fe~{\sevensize II} features, which they attributed to absorption by material from the outer accretion disc. This effect has never been directly detected before at optical wavelengths, and our optical detection is at a much higher spectral resolution than the ultraviolet \hst \ data. At these longer wavelengths we find the spectra to be dominated by Fe~{\sevensize I} features, whereas \citet{Horne94} found the majority of the absorption features in their data to be due to Fe~{\sevensize II} lines. The same effect could explain the absorption cores in the Balmer lines, which are often observed to drop below the level of the continuum in eclipsing CVs.

Using the rest wavelengths listed in Table \ref{tab:ironlines}, we fit each spectrum individually in the same way as we did for the donor star in Section \ref{sec:nir}. We plot the resulting velocities in Figure \ref{fig:uvb_indiv}. The sinusoid we plot for comparison in this figure has an amplitude of $90$km/s and a systemic velocity consistent with that derived for the donor star, and is an adequate fit to most of the velocities. There is a significant departure from this sinusoidal behaviour for the few points prior to the eclipse. This is in line with a disc origin for these absorption features since at these phases the line-of-sight will be through the bright spot, and so our measurements will be distorted by the complex  dynamics of the disc-stream interaction region. As well as the velocity, we note that the lines seem to vary in strength with phase: Figure \ref{fig:uvb_trail} shows them to be fainter (and also slightly broader) in the few spectra leading up to phase $0.5$ ($N=8$) and phase $0$ ($N=19$), but much stronger in the intermediate phases.

After removing the orbital velocity shifts and averaging the spectra, we fitted a subset of the iron lines and determined their full-width half maximum (FWHM) to be $85.5 \pm 2.9$ km/s. For this subset we excluded lines which were potentially blended. We find no significant difference between the FWHM of the lines we identify as Fe~{\sevensize I} compared with those we identify as Fe~{\sevensize II}.

\begin{figure}
\centering
\includegraphics[angle=0,width=1.0\columnwidth]{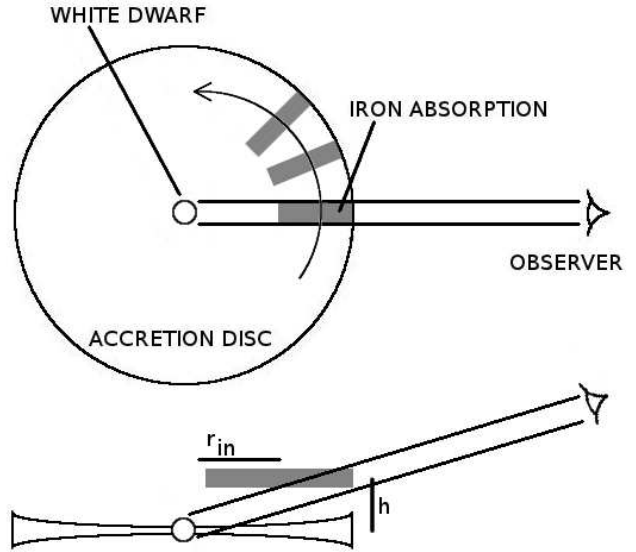}
\hfill
\caption{Cartoon showing the location of the iron absorption region. The white dwarf and the disc are not drawn to scale. The absorption region is in the outer disc and elevated at a height $h$ above the plane of the disc, so as to lie in the line-of-sight between the observer and the white dwarf. The intersection of the line-of-sight and the gas layer is at a radial distance $r_{in}$ from the white dwarf. The region of the gas layer which we observe is dependent upon the phase of the observation.}
\label{fig:cartoon} 
\end{figure}

\begin{figure}
\centering
\includegraphics[angle=270,width=1.0\columnwidth]{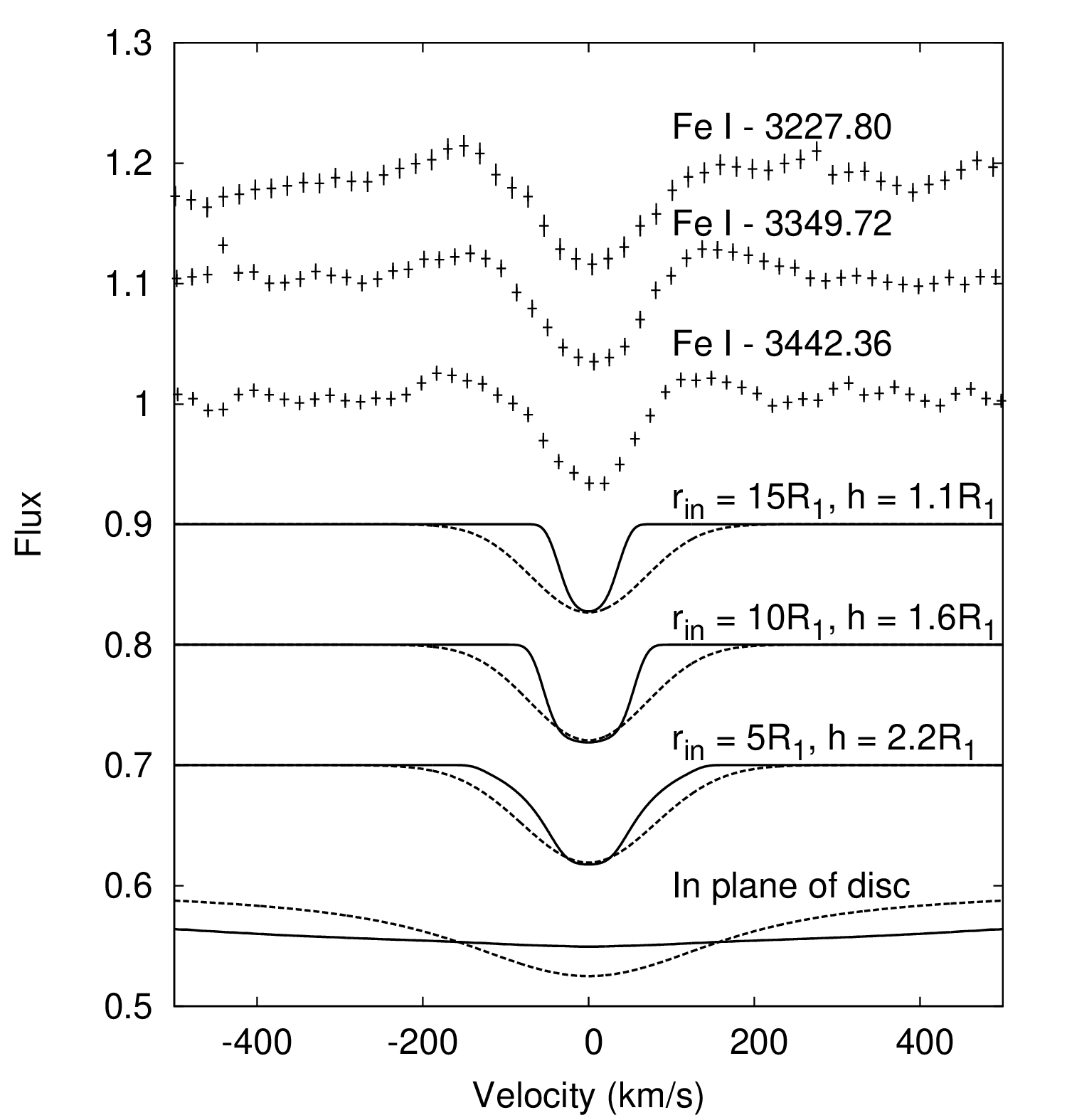}
\hfill
\caption{We show here three representative Fe lines from the iron curtain, plotting the averaged line in velocity space, with the orbital velocity variations removed. Offset below these, we show theoretical line profiles generated using the simple gas layer model described in Section \ref{sec:absmod}. The models are offset for clarity. We express all distances in terms of the white dwarf radius $R_1$ (0.01\Rsun, \citealt{Littlefair08}). We plot three models in which the absorbing layer is above the plane of the white dwarf. We choose the height in these three layers so that the minimum radius of the {\it observed} region of the gas layer $r_{in}$ is at a distance of $5$, $10$ and $15$ white dwarf radii from the white dwarf itself. We also plot a model in which the gas layer is in the plane of the white dwarf, in which case all of the absorbed light originates in the lower hemisphere of the white dwarf, and the absorption occurs in the inner disc. Finally, for each model we plot two lines, one for an assumed RMS line width of $10$km/s (solid), and on for an assumed width of $60$km/s (dashed).}
\label{fig:wabs} 
\end{figure}

\subsection{A simple model for the iron curtain lines}
\label{sec:absmod}

We computed a simple model which determines the absorption for a thin gas layer which rotates with the disc. This gas layer is assumed to be uniform in height, and is split into multiple sub-layers so the effect of shear can be accounted for. We use the binary parameters given in \citet{Littlefair08} for these calculations. We considered various locations for the absorbing gas layer. Firstly, we assumed it was very close to the white dwarf, and in the plane of the disc itself. In this configuration, the light which is absorbed originates in the lower hemisphere of the white dwarf and passes through the region of the (inclined) accretion disc which is on the line-of-sight to the observer. However, in this configuration our model predicts much broader absorption lines than we observe. In order for us to obtain line widths that match the observation, we require the observing region to be located in the outer accretion disc. Since the binary is inclined, this also requires the absorbing region to be elevated above the plane of the accretion disc, otherwise it would not intersect the light emitted from the white dwarf. In Figure \ref{fig:cartoon} we plot a cartoon which illustrates the small region of the disc we are probing. We define a height $h$ above the accretion disc for the gas layer. The $h$ we select also determines a minimum distance $r_{in}$ for the {\it observed} region of the gas layer. Note that the gas layer is not necessarily truncated at $r_{in}$, but any material closer in does not lie on the line-of-sight to the white dwarf and so is not observed. We do truncate the gas layer at the outer radius of the accretion disc, which we assume to be $\sim$$20$ white dwarf radii. Note also that as the binary phase changes the azimuth of the region we probe also changes, so if we obtain an entire orbit of time series spectra we map out an annulus of material around the disc.

In Figure \ref{fig:wabs} we plot the region around three representative iron absorption lines in the averaged spectrum. We choose lines which are strong and relatively isolated. Some of the lines show possible emission wings as well as the absorption line: if this emission is real it presumably originates in the material which is not on the line-of-sight to the white dwarf. We also plot a series of model predictions. We examine first the solid lines. For all of these models we assume an RMS line width of $10$km/s. We consider this an appropriate value to account for the turbulence in the disc. We plot the model in which the absorption is in the disc plane, as well as three models in which the gas layer is above the plane of the disc. We choose the heights of these three models such that $r_{in}$ is $5$, $10$ and $15$ white dwarf radii from the disc centre. We see that when the absorption region is in the plane of the disc the model line profile is much broader than is observed. If we consider the models in which the absorption region is in the outer disc, we see that as the distance of the absorbing region from the white dwarf increases the profile of the line narrows, and loses the broad wings which are observed at smaller radii. The models in which $r_{in}$ is between $5$ and $10$ white dwarf radii are a good match to the observed line, suggesting the iron curtain material is at a height $h$ of $1.6$ to $2.2$ white dwarf radii above the disc plane.

\citet{Horne94} attributed a velocity dispersion of $60$km/s to the iron curtain material they detected in the ultraviolet. In Figure \ref{fig:wabs} the dashed lines show the same four models when we assume the RMS line width to be equal to this velocity. The effect is to broaden the line profiles considerably. Many of the lines seem somewhat narrower than all of these models, and so we do not think the velocity dispersion of the material we observe is this high. The velocity dispersion may however be somewhere between this value and our initial assumption of $10$km/s, which is a realistic minimum. This would imply a higher $r_{in}$, which in turn would place the material closer in height to the plane of the disc than we determined for the $10$km/s models.

\subsection{The shape of the accretion disc} 
\label{sec:discshape}

\begin{figure*}
\includegraphics[angle=270,width=0.9\textwidth]{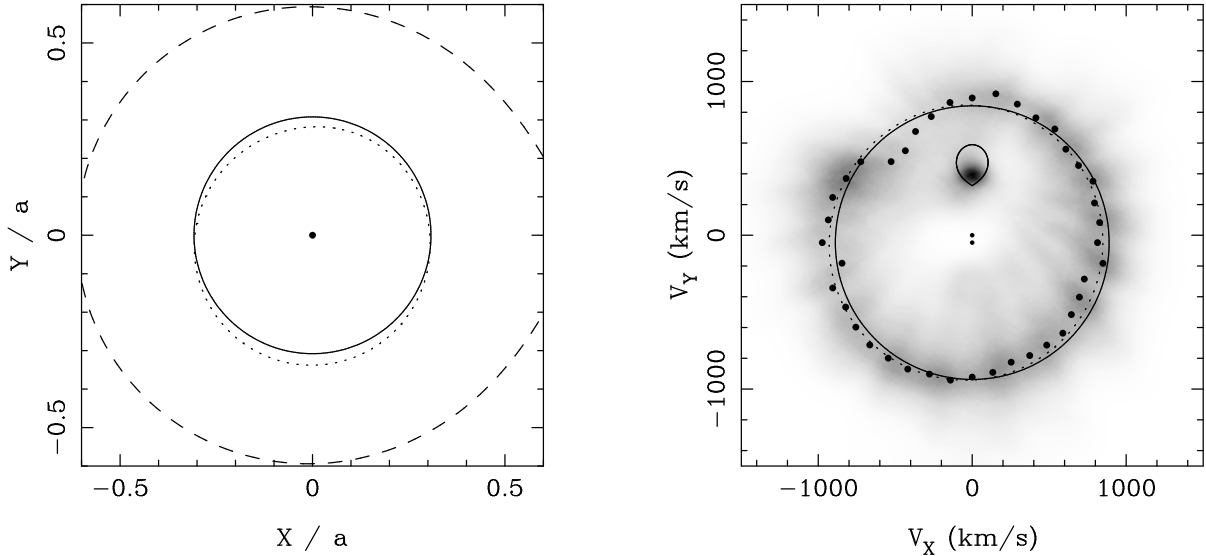}
\caption{{\bf Left:} The shape of the accretion disc, determined by fitting the iron curtain lines. The $x$ and $y$ coordinates are scaled by the binary separation $a$. When compared to a circle centred on the centre of mass(solid line), there is an distortion of the outer disc (dotted line) with the main asymmetry being in the $y$-direction. The dashed line shows the extent of the Roche lobe. {\bf Right:} Here we transform the positional information from the iron curtain fit into velocity coordinates. Again, the dotted line shows the distortion of the outer disc as determined from the iron lines. We plot also the Ca~{\sevensize II} doppler map, and the large dots denote the maxima of the emission ring. These maxima appear to track the iron curtain velocities more closely than the symmetrical circular path (solid). The two points in the centre of the plot show the centres of mass of the system and the white dwarf.} \label{fig:distortion} \end{figure*}
 
As Figure \ref{fig:uvb_indiv} shows, the radial velocities of the iron curtain do not track the velocity of the white dwarf, which we can predict with confidence given OY Car's well-determined parameters and ephemeris. The difference between the iron curtain and white dwarf velocities is evidence of significant radial motion of the material causing the absorption. For instance at phase $0$, we see material receding from us at $150$km/s while the white dwarf is moving at $66$km/s away from us (the systemic velocity). This outer disc material, which is on our line of sight to the white dwarf, thus has a radial component of $84$km/s towards the white dwarf. Taking this reasoning further, if we assume that the material we see at all phases is on a single path around the white dwarf, we can integrate the radial velocity to find its radius as a function of azimuth using
\[ R = R_0 + \int_0^t \frac{V_R}{\sin i} \, dt',\]
where $R_0$ is an arbitrary initial radius. The $\sin i$ projection factor allows for our assumption that the motion we see is parallel to the orbital plane. We can convert this into an integral over azimuthal angle $\theta$ using $\omega = d\theta/dt$, whereby
\[ R = R_0 + \int_0^\theta \frac{V_R}{\omega \sin i} \, d\theta' .\]
We take $\omega$ to be given by the standard Keplerian relation for a fixed radius $R_0$. This ignores variations that must occur given the varying radius, but should be good to first order. We select $R_0$ to match the observed outer disk velocity. The azimuthal angle is directly related to the phase $\phi$ of Figure \ref{fig:uvb_indiv} by $\theta = - 2\pi \phi$.

To enable the integration, we fitted the velocities with four Fourier components. The result is plotted in the left-hand panel of Figure \ref{fig:distortion}. A small but significant distortion of the outer disc is evident with the main asymmetry being in the $y$-direction. This stands in contrast to three-body orbits such as those of \citet{Paczynski77} which are asymmetrical in $x$ but not $y$. If the distortion deduced from the iron curtain is real, then we might expect to see it reflected in the Doppler maps, assuming that the emission lines originate from associated gas. We therefore also predicted the equivalent path in the Doppler image shown in the right-hand side of Figure \ref{fig:distortion} (dotted line). Due to the $v \propto R^{-1/2}$ relation between orbital speed and radius, this is even less distorted than the real-space figure, but we believe that it is in rough accord with the image, although the latter appears more distorted. We show this by plotting points marking the maximum of the emission ring. It can be seen that at most azimuths these points lie closer to the dotted, distorted path than they do to the solid, symmetrical circular path. We emphasize that the agreement here is between two independent lines of evidence: the path of the outer disc deduced from measurements of the iron curtain material as it is seen in absorption against the white dwarf versus the emission ring seen in the Ca~{\sevensize II}. The disc is shown to be non-circular, and the distortion although small, is enough to render any measurement of $K_1$ from the emission lines entirely useless, as is already strongly suggested by Figure \ref{fig:uvb_indiv}. It will be interesting to establish whether hydrodynamic simulations can match our observations. We cannot tell given our single orbit whether or not the distortion we see is fixed in the rotating frame of the binary or not. Further epochs of observations will be needed to determine this. This distortion could, for example be the disc ellipticity which is observed as superhumps in the light curves of SU~UMa-type CVs (such as OY~Car) in the aftermath of superoutbursts in these systems \citep{Hessman92}. A superoutburst was observed in OY~Car from 2009 November~7 to November~21\footnote{Determined from data obtained by the American Association of Variable Star Observers (http://www.aavso.org/)}, three months prior to our X-shooter observations.

\section{CONCLUSIONS}
\label{sec:discussion}

In this paper we have presented high resolution spectra of the eclipsing cataclysmic variable OY Car. The primary motivation of this work was to obtain a spectroscopic determination of the donor star orbital velocity, to compare with the value which has been inferred from photometric models of the eclipse features ($K_2 = 470 \pm 7$km/s). We find two K~{\sevensize I}
doublets in the $J$-band track the motion of the donor star, and subsequently determine $K_2 = 470.0 \pm 2.7$km/s, which is consistent with the photometric determination. As well as the absorption features from the donor star and the usual emission features seen in cataclysmic variables, we also detect a forest of absorption lines concentrated in the $3100$ -- $3500$\AA \ region. We identify the the majority of these features to be Fe~{\sevensize I} and Fe~{\sevensize II} lines, and they have an approximately sinusoidal radial velocity curve with a phase and amplitude which is not consistent with an white dwarf or donor star origin. We attribute these features to the `iron curtain' of absorbing material from the outer disc, which veils the white dwarf in this system and was first reported by \citet{Horne94}. The velocities of these lines do not track the white dwarf, reflecting a distortion of the outer disc that we see also in the Doppler images. This is evidence for considerable radial motion in the outer disk, at up to $90$km/s towards and away from the white dwarf.

\section*{ACKNOWLEDGEMENTS}
CMC, TRM, VSD and EB are supported by rolling grants from the Science and Technology Facilities Council (STFC). DS acknowledges the support of an STFC
Advanced Fellowship. The results presented in this paper are based on observations collected at ESO, Chile (Programme 084.D-1149). We also acknowledge with thanks the variable star observations from the AAVSO International Database contributed by observers worldwide and used in this research. This research has made use of NASA's Astrophysics Data System Bibliographic Services and the SIMBAD data base, operated at CDS, Strasbourg, France. We thank the reviewer Knox Long for his useful comments.

\bibliography{oycar}

\begin{thebibliography}{}

\bibitem[\protect\citeauthoryear{{Bailey}}{{Bailey}}{1979}]{Bailey79}
{Bailey} J.,  1979, \mnras, 187, 645

\bibitem[\protect\citeauthoryear{{Billington}, {Marsh} \&
  {Dhillon}}{{Billington} et~al.}{1996}]{Billington96}
{Billington} I.,  {Marsh} T.~R.,    {Dhillon} V.~S.,  1996, \mnras, 278, 673

\bibitem[\protect\citeauthoryear{{Claret}}{{Claret}}{1998}]{Claret98}
{Claret} A.,  1998, \aap, 335, 647

\bibitem[\protect\citeauthoryear{{Cook} \& {Warner}}{{Cook} \&
  {Warner}}{1984}]{Cook84}
{Cook} M.~C.,  {Warner} B.,  1984, \mnras, 207, 705

\bibitem[\protect\citeauthoryear{{Copperwheat}, {Marsh}, {Dhillon},
  {Littlefair}, {Hickman}, {G{\"a}nsicke} \& {Southworth}}{{Copperwheat}
  et~al.}{2010}]{Copperwheat10}
{Copperwheat} C.~M.,  {Marsh} T.~R.,  {Dhillon} V.~S.,  {Littlefair} S.~P.,
  {Hickman} R.,  {G{\"a}nsicke} B.~T.,    {Southworth} J.,  2010, \mnras, 402,
  1824

\bibitem[\protect\citeauthoryear{{Copperwheat}, {Marsh}, {Littlefair},
  {Dhillon}, {Ramsay}, {Drake}, {G{\"a}nsicke}, {Groot}, {Hakala}, {Koester},
  {Nelemans}, {Roelofs}, {Southworth}, {Steeghs} \& {Tulloch}}{{Copperwheat}
  et~al.}{2011}]{Copperwheat11}
{Copperwheat} C.~M.,  {Marsh} T.~R.,  {Littlefair} S.~P.,  {Dhillon} V.~S.,
  {Ramsay} G.,  {Drake} A.~J.,  {G{\"a}nsicke} B.~T.,  {Groot} P.~J.,  {Hakala}
  P.,  {Koester} D.,  {Nelemans} G.,  {Roelofs} G.,  {Southworth} J.,
  {Steeghs} D.,    {Tulloch} S.,  2011, \mnras, 410, 1113

\bibitem[\protect\citeauthoryear{{Cushing}, {Rayner} \& {Vacca}}{{Cushing}
  et~al.}{2005}]{Cushing05}
{Cushing} M.~C.,  {Rayner} J.~T.,    {Vacca} W.~D.,  2005, \apj, 623, 1115

\bibitem[\protect\citeauthoryear{{D'Odorico}, {Dekker}, {Mazzoleni}, {Vernet},
  {Guinouard}, {Groot}, {Hammer}, {Rasmussen}, {Kaper}, {Navarro},
  {Pallavicini}, {Peroux} \& {Zerbi}}{{D'Odorico} et~al.}{2006}]{Dodorico06}
{D'Odorico} S.,  {Dekker} H.,  {Mazzoleni} R.,  {Vernet} J.,  {Guinouard} I.,
  {Groot} P.,  {Hammer} F.,  {Rasmussen} P.~K.,  {Kaper} L.,  {Navarro} R.,
  {Pallavicini} R.,  {Peroux} C.,    {Zerbi} F.~M.,  2006, in Society of
  Photo-Optical Instrumentation Engineers (SPIE) Conference Series Vol.~6269 of
  Society of Photo-Optical Instrumentation Engineers (SPIE) Conference Series,
  {X-shooter UV- to K-band intermediate-resolution high-efficiency spectrograph
  for the VLT: status report at the final design review}

\bibitem[\protect\citeauthoryear{{Efron}}{{Efron}}{1979}]{Efron79}
{Efron} B.,  1979, Annals of Statistics, 7, 1

\bibitem[\protect\citeauthoryear{{Efron} \& {Tibshirani}}{{Efron} \&
  {Tibshirani}}{1993}]{Efron93}
{Efron} B.,  {Tibshirani} R.~J.,  1993, {An Introduction to the Bootstrap}.
Chapman \& Hall, New York

\bibitem[\protect\citeauthoryear{{Eggleton}}{{Eggleton}}{1983}]{Eggleton83}
{Eggleton} P.~P.,  1983, \apj, 268, 368

\bibitem[\protect\citeauthoryear{{Greenhill}, {Hill}, {Dieters}, {Fienberg},
  {Howlett}, {Meijers}, {Munro} \& {Senkbeil}}{{Greenhill}
  et~al.}{2006}]{Greenhill06}
{Greenhill} J.~G.,  {Hill} K.~M.,  {Dieters} S.,  {Fienberg} K.,  {Howlett} M.,
   {Meijers} A.,  {Munro} A.,    {Senkbeil} C.,  2006, \mnras, 372, 1129

\bibitem[\protect\citeauthoryear{{Hessman}, {Mantel}, {Barwig} \&
  {Schoembs}}{{Hessman} et~al.}{1992}]{Hessman92}
{Hessman} F.~V.,  {Mantel} K.-H.,  {Barwig} H.,    {Schoembs} R.,  1992, \aap,
  263, 147

\bibitem[\protect\citeauthoryear{{Horne}, {Marsh}, {Cheng}, {Hubeny} \&
  {Lanz}}{{Horne} et~al.}{1994}]{Horne94}
{Horne} K.,  {Marsh} T.~R.,  {Cheng} F.~H.,  {Hubeny} I.,    {Lanz} T.,  1994,
  \apj, 426, 294

\bibitem[\protect\citeauthoryear{{Horne}, {Wade} \& {Szkody}}{{Horne}
  et~al.}{1986}]{Horne86}
{Horne} K.,  {Wade} R.~A.,    {Szkody} P.,  1986, \mnras, 219, 791

\bibitem[\protect\citeauthoryear{{Kaitchuck}, {Schlegel}, {Honeycutt}, {Horne},
  {Marsh}, {White} II \& {Mansperger}}{{Kaitchuck} et~al.}{1994}]{Kaitchuck94}
{Kaitchuck} R.~H.,  {Schlegel} E.~M.,  {Honeycutt} R.~K.,  {Horne} K.,  {Marsh}
  T.~R.,  {White} II J.~C.,    {Mansperger} C.~S.,  1994, \apjs, 93, 519

\bibitem[\protect\citeauthoryear{{Littlefair}, {Dhillon}, {Howell} \&
  {Ciardi}}{{Littlefair} et~al.}{2000}]{Littlefair00}
{Littlefair} S.~P.,  {Dhillon} V.~S.,  {Howell} S.~B.,    {Ciardi} D.~R.,
  2000, \mnras, 313, 117

\bibitem[\protect\citeauthoryear{{Littlefair}, {Dhillon}, {Marsh},
  {G{\"a}nsicke}, {Southworth}, {Baraffe}, {Watson} \&
  {Copperwheat}}{{Littlefair} et~al.}{2008}]{Littlefair08}
{Littlefair} S.~P.,  {Dhillon} V.~S.,  {Marsh} T.~R.,  {G{\"a}nsicke} B.~T.,
  {Southworth} J.,  {Baraffe} I.,  {Watson} C.~A.,    {Copperwheat} C.,  2008,
  \mnras, 388, 1582

\bibitem[\protect\citeauthoryear{{Marsh}}{{Marsh}}{1988}]{Marsh88a}
{Marsh} T.~R.,  1988, \mnras, 231, 1117

\bibitem[\protect\citeauthoryear{{Marsh}}{{Marsh}}{2001}]{Marsh01}
{Marsh} T.~R.,  2001, in {H.~M.~J.~Boffin, D.~Steeghs, \& J.~Cuypers} ed.,
  Astrotomography, Indirect Imaging Methods in Observational Astronomy Vol.~573
  of Lecture Notes in Physics, Berlin Springer Verlag, {Doppler Tomography}.
pp~1--+

\bibitem[\protect\citeauthoryear{{Marsh} \& {Horne}}{{Marsh} \&
  {Horne}}{1988}]{Marsh88}
{Marsh} T.~R.,  {Horne} K.,  1988, \mnras, 235, 269

\bibitem[\protect\citeauthoryear{{Marsh}, {Horne}, {Schlegel}, {Honeycutt} \&
  {Kaitchuck}}{{Marsh} et~al.}{1990}]{Marsh90}
{Marsh} T.~R.,  {Horne} K.,  {Schlegel} E.~M.,  {Honeycutt} R.~K.,
  {Kaitchuck} R.~H.,  1990, \apj, 364, 637

\bibitem[\protect\citeauthoryear{{Marsh}, {Robinson} \& {Wood}}{{Marsh}
  et~al.}{1994}]{Marsh94}
{Marsh} T.~R.,  {Robinson} E.~L.,    {Wood} J.~H.,  1994, \mnras, 266, 137

\bibitem[\protect\citeauthoryear{{Nelemans}, {Yungelson} \& {Portegies
  Zwart}}{{Nelemans} et~al.}{2004}]{Nelemans04}
{Nelemans} G.,  {Yungelson} L.~R.,    {Portegies Zwart} S.~F.,  2004, \mnras,
  349, 181

\bibitem[\protect\citeauthoryear{{Paczynski}}{{Paczynski}}{1977}]{Paczynski77}
{Paczynski} B.,  1977, \apj, 216, 822

\bibitem[\protect\citeauthoryear{{Press}}{{Press}}{2002}]{Press02}
{Press} W.~H.,  2002, {Numerical recipes in C++ : the art of scientific
  computing}.
Cambridge University Press

\bibitem[\protect\citeauthoryear{{Ralchenko}, {Kramida}, Reader \& {NIST ASD
  Team (2010)}}{{Ralchenko} et~al.}{2010}]{Ralchenko11}
{Ralchenko} Y.,  {Kramida} A.~E.,  Reader J.,    {NIST ASD Team (2010)} 2010,
  {NIST Atomic Spectra Database (version 4.0), [Online]. Available:
  http://physics.nist.gov/asd}.
National Institute of Standards and Technology, Gaithersburg, MD.

\bibitem[\protect\citeauthoryear{{Savoury}, {Littlefair}, {Dhillon}, {Marsh},
  {Copperwheat} \& {Parsons}}{{Savoury} et~al.}{2011}]{Savoury11a}
{Savoury} C.~D.~J.,  {Littlefair} S.~P.,  {Dhillon} V.~S.,  {Marsh} T.~R.,
  {Copperwheat} C.~M.,    {Parsons} S.~G.,  2011, {\mnras (submitted)}, p.~1

\bibitem[\protect\citeauthoryear{{Savoury}, {Littlefair}, {Dhillon}, {Marsh},
  {G{\"a}nsicke}, {Copperwheat}, {Kerry}, {Hickman} \& {Parsons}}{{Savoury}
  et~al.}{2011}]{Savoury11}
{Savoury} C.~D.~J.,  {Littlefair} S.~P.,  {Dhillon} V.~S.,  {Marsh} T.~R.,
  {G{\"a}nsicke} B.~T.,  {Copperwheat} C.~M.,  {Kerry} P.,  {Hickman} R.~D.~G.,
     {Parsons} S.~G.,  2011, \mnras, pp 1050--+

\bibitem[\protect\citeauthoryear{{Smak}}{{Smak}}{1979}]{Smak79}
{Smak} J.,  1979, \actaa, 29, 309

\bibitem[\protect\citeauthoryear{{Szkody}, {Anderson}, {Hayden}, {Kronberg},
  {McGurk}, {Riecken}, {Schmidt}, {West}, {G{\"a}nsicke}, {Nebot Gomez-Moran},
  {Schneider}, {Schreiber} \& {Schwope}}{{Szkody} et~al.}{2009}]{Szkody09}
{Szkody} P.,  {Anderson} S.~F.,  {Hayden} M.,  {Kronberg} M.,  {McGurk} R.,
  {Riecken} T.,  {Schmidt} G.~D.,  {West} A.~A.,  {G{\"a}nsicke} B.~T.,  {Nebot
  Gomez-Moran} A.,  {Schneider} D.~P.,  {Schreiber} M.~R.,    {Schwope} A.~D.,
  2009, \aj, 137, 4011

\bibitem[\protect\citeauthoryear{{Tulloch}, {Rodr{\'{\i}}guez-Gil} \&
  {Dhillon}}{{Tulloch} et~al.}{2009}]{Tulloch09}
{Tulloch} S.~M.,  {Rodr{\'{\i}}guez-Gil} P.,    {Dhillon} V.~S.,  2009, \mnras,
  397, L82

\bibitem[\protect\citeauthoryear{{Wade} \& {Horne}}{{Wade} \&
  {Horne}}{1988}]{Wade88}
{Wade} R.~A.,  {Horne} K.,  1988, \apj, 324, 411

\bibitem[\protect\citeauthoryear{{Warner}}{{Warner}}{1995}]{Warner95}
{Warner} B.,  1995, Cambridge Astrophysics Series, 28

\bibitem[\protect\citeauthoryear{{Watson}, {Dhillon}, {Rutten} \&
  {Schwope}}{{Watson} et~al.}{2003}]{Watson03}
{Watson} C.~A.,  {Dhillon} V.~S.,  {Rutten} R.~G.~M.,    {Schwope} A.~D.,
  2003, \mnras, 341, 129

\bibitem[\protect\citeauthoryear{{Wood}, {Horne}, {Berriman}, {Wade},
  {O'Donoghue} \& {Warner}}{{Wood} et~al.}{1986}]{Wood86}
{Wood} J.,  {Horne} K.,  {Berriman} G.,  {Wade} R.,  {O'Donoghue} D.,
  {Warner} B.,  1986, \mnras, 219, 629

\bibitem[\protect\citeauthoryear{{Wood} \& {Horne}}{{Wood} \&
  {Horne}}{1990}]{Wood90}
{Wood} J.~H.,  {Horne} K.,  1990, \mnras, 242, 606

\end{thebibliography}

\end{document}